\documentclass{aa}  

\usepackage{graphicx}

\usepackage{txfonts}
\begin{document}

   \title{Modelling shock-like injections of solar energetic particles with 3D test particle simulations}

   \subtitle{}

   \author{A. Hutchinson
          \inst{1}
          \and
          S. Dalla\inst{1}
          \and
          T. Laitinen\inst{1}
          \and
          C. O. G. Waterfall\inst{1}}

   \institute{Jeremiah Horrocks Institute, University of Central Lancashire, Preston, PR1 2HE, UK\\
                \email{AHutchinson3@uclan.ac.uk}}
             

 
  \abstract
   {Solar Energetic Particle (SEP) acceleration and injection into interplanetary space during gradual SEP events is thought to take place at Coronal Mass Ejection (CME)-driven shocks. Various features of measured intensity profiles at 1 au have been attributed to properties of the radial and longitudinal/latitudinal SEP injections at the shock. Focussed transport models are typically used to model acceleration at a CME-shock and subsequent propagation. Test particle simulations are an alternative approach but so far they have been carried out only with instantaneous injection near the Sun. }
   {We develop the first temporally extended shock-like injection for our 3D test particle code and investigate how the spatial features of injection at a shock affect SEP intensity and anisotropy profiles for observers at 0.3 and 1.0 au.}
   {We conduct simulations of a monoenergetic population of 5 MeV protons considering three different radial injection functions and two longitudinal/latitudinal injection functions. We consider a range of scattering conditions with scattering mean free path values ranging from $\lambda = 0.1 - 1.0$ au, and determine intensity and anisotropy profiles at six observers at different longitudinal locations.}
   {We find that the radial, longitudinal and latitudinal injection functions play a relatively minor role in shaping the SEP intensity profiles. The dependence of intensity profiles on the value of the scattering mean free path is also weak, unlike what is found from 1D focussed transport models. Spatial factors, such as the time of observer-shock-connection/disconnection and time of shock passage have a much stronger influence on SEP intensities and anisotropies. Persistent anisotropies until shock passage are seen in our simulations.
   Comparing instantaneous and shock-like injections, we find that the link between duration of injection and duration of the SEP event is very weak, unlike what is commonly assumed.  }
  {} 


   \maketitle
%

\section{Introduction}
Gradual Solar Energetic Particle (SEP) events are enhancements in energetic ions and electrons in the interplanetary medium, observed in association with solar flares and Coronal Mass Ejections (CMEs). They are often found to have a long duration (e.g. several days) and wide longitudinal extent (e.g. $> 80^\circ$), as evidenced by multi-spacecraft observations \citep[e.g.][and references therein.]{Desai_2016}.

Gradual events are thought to originate at CME-driven shocks, with particle acceleration taking place over extended times as the shock propagates though the corona and interplanetary space.
Several properties of the SEP intensity profiles have been linked to properties of the shock acceleration. It is thought that the characteristic long duration decays measured by spacecraft at 1 au \citep{cane_1988} are caused by the fact that CME-shock acceleration is extended in time \citep{Reames_1997}. In addition the dependence of the overall shape of SEP intensity profiles on the location of the associated solar active region, AR, \citep[east-west effect,][]{cane_1988} has been attributed to the longitudinal variation of efficiency across the the shock front \citep[e.g.][]{Tylka_2006,Zan_2006}.

A number of models have studied SEP acceleration/injection at CME-driven shocks with the purpose of deriving observables at 1 au, and determining the main physical processes that influence them.
Several studies have modelled shock-like injections and the subsequent propagation of SEPs using particle transport equations  \citep[e.g.][]{Kal&Wib_1997,Lario_1998,He_2011,Wang_2012,Qin_2013}.

\cite{Kal&Wib_1997} used a black box shock model, and solved the 1D focussed transport equation to produce intensity and anisotropy profiles for a variety of particle injection and propagation parameters. \cite{Lario_1998} modelled SEP events with an improved transport equation that considered solar wind convection and adiabatic deceleration, and used a Magnetohydrodynamic (MHD) model to derive the shock propagation. They fit observed particle flux and anisotropy profiles for four particle events and found that the efficiency of the shock as a particle accelerator decreases rapidly with distance from the Sun for particle energies greater than $\sim 2$ MeV.
\cite{heras_1994} emphasised the importance of the geometry of magnetic connection between the shock and the observer, in determining features of SEP intensity profiles, and stressed the importance of the time-evolution of magnetic connection to the shock.



Alternative to focussed transport models, full-orbit test-particle simulations of SEP propagation calculate individual particle trajectories throughout the heliosphere by solving the particle's equation of motion. These types of simulations are intrinsically 3D. Previously, we used test particle simulations to study propagation through the interplanetary magnetic field for the case of instantaneous particle injections close to the Sun (e.g. \cite{Marsh_2015,Battarbee_2018, Dalla_2020,Waterfall_2022}). \cite{Hutch_2022} used instantaneous particle injections at a range of heights in the solar atmosphere to study proton back-precipitation to the photosphere. 

In this paper we present results from 3D test particle simulations in which, for the first time, particle injection is from a moving shock-like source. This means that it is spatially extended over a wide shock and temporally extended since particles keep being injected as the shock propagates though the heliosphere. This new injection considers a simple model of a shock, similar to that developed by \cite{heras_1994} and \cite{Kal&Wib_1997}, and provides a first step in accounting for temporally extended particle injections in 3D full orbit test particle simulations.

We investigate the effect of choosing different radial and angular injection profiles on the intensity and anisotropy profiles at 1 au at different observers. We study the effect of a variety of interplanetary scattering conditions on observables. We compare intensity and anisotropy profiles for observers at 0.3 and 1.0 au, to provide comparisons with \textit{Parker Solar Probe} and \textit{Solar Orbiter} observations.
In a companion paper \citep{hutch_2022-in_prep} a comparison between simulations that do/do not include the corotation of magnetic flux tubes is presented.
The layout of the paper is as follows: in section \ref{sec:Inj} we describe the shock-like injection. In section \ref{sec:simulation} its implementation within 3D test particle simulations. In section \ref{sec:Geometry} we describe the geometry of the observers. Intensity and anisotropy profiles are investigated for variety of shock-like injections and scattering parameters in section \ref{sec:int_profs}. Discussion and conclusions can be found in section \ref{sec:disc}.

\section{Shock-like injection model}\label{sec:Inj}

We have developed a particle injection model that approximates a temporally extended injection from a propagating shock-like source (referred to in the following as the shock), of constant angular width in longitude and latitude. The injection is spatially extended across the shock front and temporally extended, i.e. occurring over timescales of days as the shock propagates radially outward. Our simplified description accounts for the geometric characteristics of the shock injection and its evolution with time, but it does not model acceleration nor the evolution of the shock as an MHD structure. We note that the shock has negligible thickness within the model.


We use a spherical coordinate system ($r,\: \theta,  \: \phi$) where $r$ is the radial distance from the centre of the Sun, $\theta$ is heliographic colatitude and $\phi$ is heliographic longitude with corresponding unit vectors $\mathbf{\hat{e}_r}$, $\mathbf{\hat{e}_\theta}$, $\mathbf{\hat{e}_\phi}$. The shock is assumed to propagate with a constant speed, $v_{sh}$, and maintain the same angular width in longitude and latitude as it propagates. Particle injection is assumed to continue as the shock propagates through the corona and interplanetary space, taking place over $r \in$ [$r_{min}$, $r_{max}$], where $r_{min}$ and $r_{max}$ are the minimum and maximum particle injection radii. We choose $r_{min} = 1.2 \: \textrm{R}_\odot$, the median shock formation height as determined by \cite{Gop_2013}.

The shock in the model acts only as a particle source and does not interact with or modify the Interplanetary Magnetic Field (IMF), which is assumed to be a Parker spiral (Equation \ref{parker_spiral}). Particles are injected isotropically from the shock surface throughout its propagation, consistent with theories of diffusive shock acceleration \citep{Baring_1997,Desai_2016}. The shock is effectively transparent to particles, meaning any particle that returns to the shock position during its propagation does not interact with it and travels straight through, without change to its trajectory.

Table \ref{params_table} summarises the shock parameters and gives the typical values used in our simulations unless otherwise stated. We define $ w_{sh,\phi}$ as the longitudinal width of the shock, $w_{sh,\theta}$ as the colatitudinal width, and $\delta_{cent}$ as the latitude of the centre of the shock (where $\delta = 90^\circ - \theta$ is latitude). In the majority of our simulations we choose $w_{sh,\phi} = w_{sh,\theta} = 70^\circ$, purely radial shock speed $v_{sh} = 1500 $ km/s, and we assume that particle acceleration/injection takes place over a period of two days, resulting in $r_{max} =  372$ R$_\odot$ ($= 1.73$ au). 
The shock is centred at $\phi_{cent} = 0^\circ$ and latitude, $\delta_{cent} = 15^\circ$, so as to represent the typical CME latitudes, corresponding to the positions of the activity belts on the solar disc.

\begin{table}[h!]
    \caption{Parameters of shock-like injection and in our test particle simulations. Columns are from left to right: variable name, parameter description and typical value used.}
    \centering
    \begin{tabular}{ c c c }
        \hline
        {Variable} & {Description} & {Value}  \\
        \hline 
         
        $v_{sh}$ & {Radial velocity of the shock} & 1500 km/s \\
        
        $w_{sh,\phi}$ & {Longitudinal width of the shock} & $70^{\circ}$ \\
        
        $w_{sh,\theta}$ & {Latitudinal height of the shock} & $70^{\circ}$\\
        
        $\delta_{cent}$ & {Latitude of shock centre} & $15^{\circ}$ \\
        
         $\phi_{cent}$ & {Longitude of shock centre} & $0^{\circ}$ \\
        
        $\sigma_{\phi}$ & {Longitudinal $\sigma$ of injection efficiency \tablefootmark{a}} & $17.5^{\circ}$ \\
        
        $\sigma_{\theta}$ & {Latitudinal $\sigma$ of injection efficiency \tablefootmark{a} } & $17.5^{\circ}$ \\ 
        
        $r_{min}$ & {Minimum radial injection position} & 1.2 R$_{\odot}$ \\
        
        $r_{max}$ & {Maximum radial injection position} &  1.73 au \\
        
        $r_{peak}$ & {Radial position of peak injection \tablefootmark{b}} & 5 R$_{\odot}$ \\ 
        
        $N_p$ & {Total number of protons injected} & 1$\times 10^6$ \\
        \hline
    \end{tabular}
    \tablefoot{\\
    \tablefoottext{a}{Only used for Gaussian $\Phi(\phi)$ and $\Theta(\theta)$}
    \tablefoottext{b}{Only used for Weibull function injection in $r$}}
    
    \label{params_table}
\end{table}

\subsection{Particle injection ad normalisation}
We introduce a function $S(r,\theta,\phi)$ describing the number of particles per unit volume injected by the shock at position ($r$, $\theta$, $\phi$), so that 
\begin{equation}
 \int d^3\mathbf{r} \: S(r, \theta, \phi) = N_p
\end{equation}
with $N_p$ the total number of particles and $d^3\mathbf{r}$ the unit volume. Here the dependence on time is folded into the radial dependence via $r\: = \: v_{sh} \: t$.

We assume that $S$ is separable, so that;
\begin{equation}
   S(r, \theta, \phi) = N_p \:\frac{R(r)}{r^2} \: \frac{\Theta(\theta)}{\sin\theta} \: \Phi(\phi) \:  
\end{equation}

where $R(r)$, $\Phi(\phi)$, $\Theta(\theta)$ are the radial, longitudinal and colatitudinal injection functions, satisfying the normalisation conditions:

\begin{equation}
    \int dr \: R(r) = 1
\end{equation}
\begin{equation}    
    \int d\theta \: \Theta(\theta) = 1
\end{equation}
\begin{equation}    
    \int d\phi \: \Phi(\phi) = 1
\end{equation}

\subsection{Radial injection function}

In our model we consider a number of radial injection functions. The simplest radial injection function is $R(r) = const$, the uniform case, where the shock injects the same number of particles as $r$ increases. The normalisation condition for the uniform case gives:

\begin{equation}
    R(r) = \frac{1}{(r_{max} - r_{min})}.
    \label{uniform_r}
\end{equation}

\begin{figure}[h!]
    \centering
    \includegraphics[width = 0.9\linewidth, keepaspectratio = true]{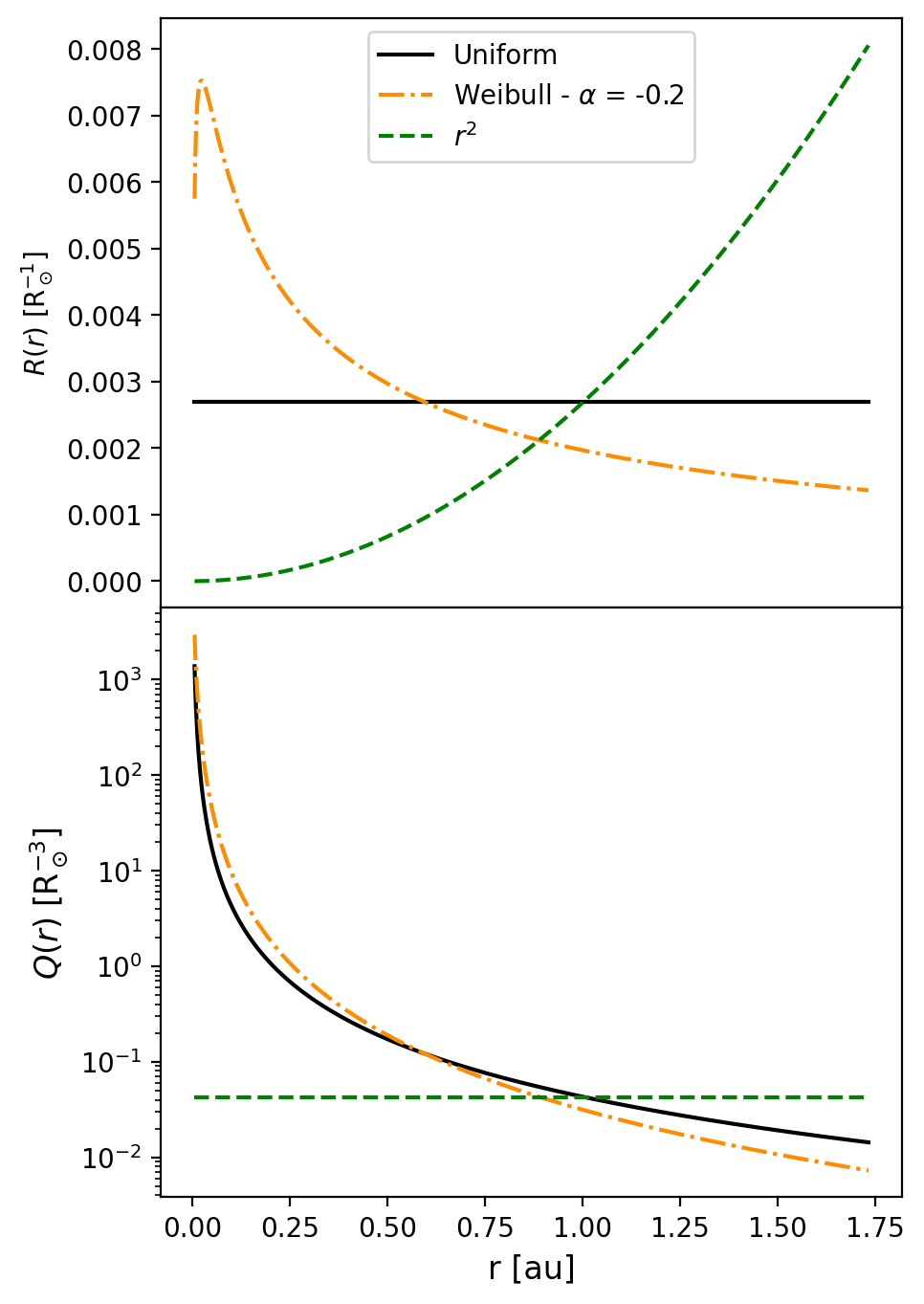}
    \caption{Radial injection function, $R(r)$ versus $r$ (top) and the number of particles injected per unit area of the shock, $Q(r)$ versus $r$ (bottom).}
    \label{Inj_rad_plt}
\end{figure}

We note that as the shock propagates to larger radial positions with fixed angular width the surface area of the shock increases as $r^2$. Therefore for uniform $R(r)$ (Equation \ref{uniform_r}) the number of particles injected per unit area of the shock, $Q(r)$, decreases with $r$ like $1/r^2$ (see Figure \ref{Inj_rad_plt} bottom panel).

We also consider a non-uniform injection given by the modified Weibull function \citep[previously used to describe SEP profiles by][]{Kahler_2018}. Normalised over our injection range this is given by,

\begin{equation}
    R(r) = \frac{(-\alpha / \beta) (r/\beta)^{\alpha - 1} \exp(-(r/\beta)^\alpha)}{\exp\left(-(r_{max}/\beta)^\alpha\right) - \exp\left(-(r_{min}/\beta)^\alpha\right)},
    \label{Weibull_func}
\end{equation}
where $\alpha$ ($\alpha < 0$) and $\beta$ are parameters that determine the shape of the function, specifically the rise and decay rates and the position of the peak. In our work we choose $r_{peak} = 5$ R$_\odot$, which results in a fast rise phase near the Sun and a decay phase beyond 5 R$_\odot$. The Weibull function is plotted in orange in Figure \ref{Inj_rad_plt} (top panel).

 To study the case of constant injection per unit area of the shock, we also consider a radial injection function proportional to $r^2$, This function is described by,

\begin{equation}
    R(r) = \frac{3r^2}{(r_{max}^3 - r_{min}^3)}.
    \label{r_squared_inj}
\end{equation}
The number of injected particles per unit area of the shock is,

\begin{equation}
    Q(r) = \frac{R(r) \: N_p}{A(r)}
\end{equation}
where $A(r)$ is the surface area of the shock front at a given radial position, given by,

\begin{equation}
    A(r) = 2 \:r^2 \: w_{sh,\phi} \: \cos(\delta_{cent}) \: \sin(w_{sh,\theta}/2) 
\end{equation}

Figure \ref{Inj_rad_plt} (top) displays all radial injection functions, $R(r)$ versus radial shock position and Figure \ref{Inj_rad_plt} (bottom) displays the corresponding $Q(r)$. 

\subsection{Longitudinal/latitudinal injection functions}

We consider two types of injection efficiency with respect to longitudinal and latitudinal position across the shock front. The first is uniform injection efficiency across the shock front, described by the normalised expressions:

\begin{equation}
    \Phi(\phi) = \frac{1}{ w_{sh,\phi}}
    \label{uniform_long}
\end{equation}

\begin{equation}
    \Theta(\theta) = \frac{1}{2 \cos(\delta_{cent}) \sin( w_{sh,\theta}/2)}
    \label{uniform_long}
\end{equation}

\begin{figure}
    \centering
    \includegraphics[width = 1.0\linewidth, keepaspectratio = true]{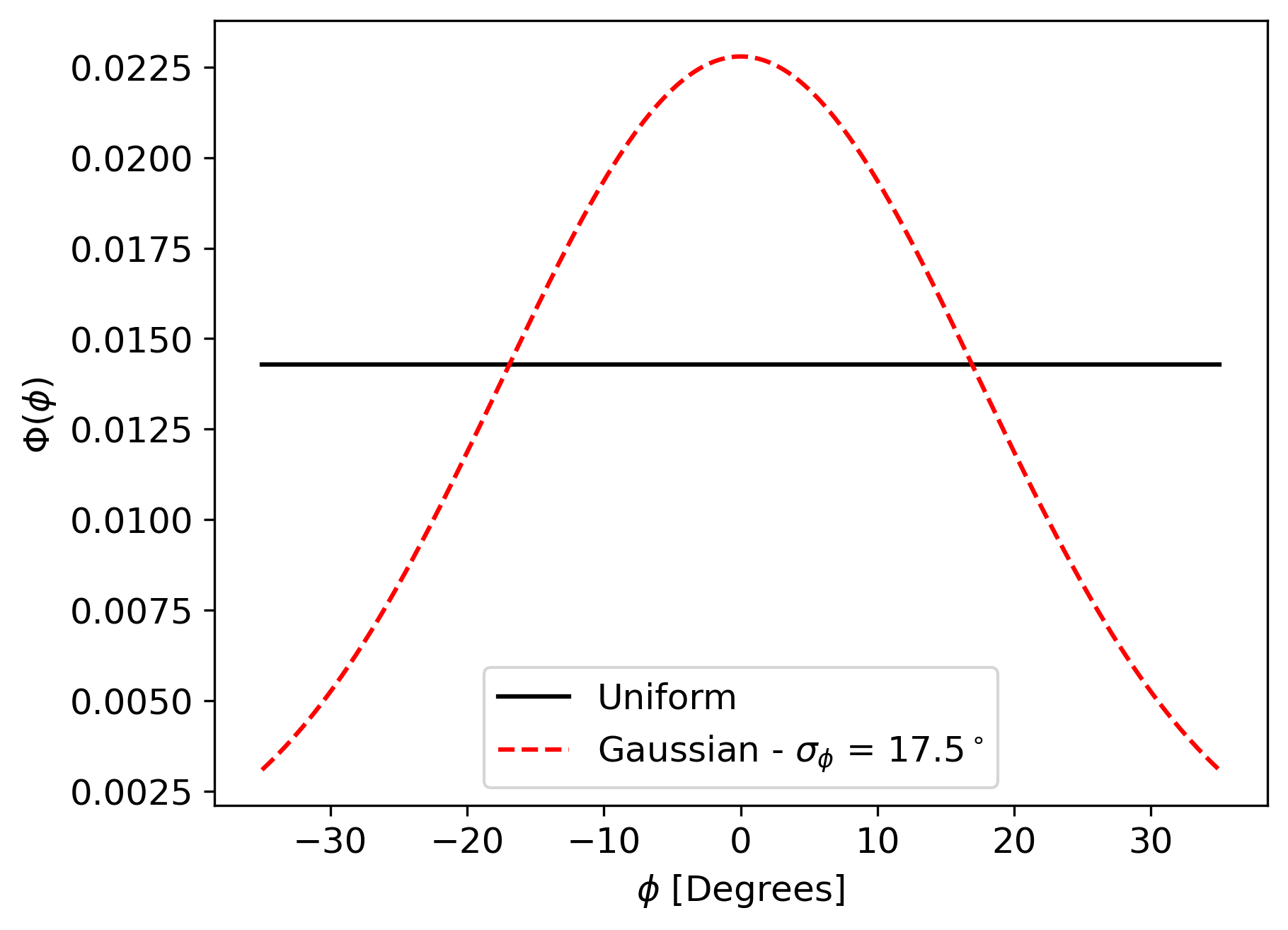}
    \caption{Longitudinal injection function, $\Phi(\phi)$ for the shock-like injection. }
    \label{long_func}
\end{figure}

The second longitudinal and latitudinal injection function is a Gaussian centred about the shock nose. Typically we consider standard deviations of $\sigma_{\phi} \: (\sigma_{\theta}) =  w_{sh,\phi}/4 \: ( w_{sh,\theta}/4)$. The two injection functions are displayed in Figure \ref{long_func} for the longitudinal injection efficiency. 

\section{3D full-orbit test particle simulations}\label{sec:simulation}

We model SEP propagation using a 3D full-orbit test particle code, originally developed by \cite{Dalla&Browning_2005}, that has since been adapted to model various aspects of SEP propagation \citep[e.g.][]{Marsh_2013,Battarbee_2018}.

The model calculates particle trajectories by integrating their equation of motion,

\begin{equation}
    \frac{d\mathbf{p}}{dt} = q \left( \mathbf{E} + \frac{1}{c}\frac{\mathbf{p}}{m_0 \gamma}\times\mathbf{B} \right)
    \label{eqn_o_motion}
\end{equation}
where $\mathbf{p}$ is the particle momentum, $c$ is the speed of light, $t$ is time, $q$ is the charge of the particle, $m_0$ is the particle rest mass, $\gamma$ is the Lorentz factor and $\mathbf{E}$ and $\mathbf{B}$ are the electric and magnetic fields respectively. 

Within this model we consider a Parker spiral magnetic field described by,

\begin{equation}
    \mathbf{B} = \frac{B_0 r_0^2}{r^2} \mathbf{\hat{e}_r} - \frac{B_0 r_0^2 \Omega \sin{\theta}}{v_{sw} r} \mathbf{\hat{e}_\phi},
    \label{parker_spiral}
\end{equation}
where $B_0$ is the magnetic field strength at a fixed reference radial distance $r_0$, $\Omega$ is the sidereal solar rotation rate, and $v_{sw}$ is the solar wind speed. In our simulations we consider a unipolar positive magnetic field (i.e. antisunward oriented) and solar wind speed of 500 km s$^{-1}$ \citep[same parameter values as][]{Marsh_2013}.
An electric field is present in the non-rotating (spacecraft) coordinate system due to the outward flow of the solar wind, and is defined by \citep[][]{Marsh_2013,dalla_2013}:
\begin{equation}
    \mathbf{E} = \frac{-B_0 \: r_0^2 \:\Omega \: \sin\theta}{c \: r} \mathbf{\hat{e}_\theta}
    \label{sol_win_E}
\end{equation}
Its effect on the particle trajectories is to make particles corotate with the magnetic flux tubes in which they propagate. Hence corotation effects are included in our simulations.
Within the model we consider the effects of magnetic turbulence by introducing pitch angle scattering, characterised by a mean free path $\lambda$. For a further details of the scattering model see \cite{Marsh_2013} and \cite{Dalla_2020}.

We note that in this first study of the effects of time extended injection, our model does not include the effects of turbulence-associated perpendicular diffusion, which would effect the intensity profiles observed at different longitudinal positions, as shown by \cite{Wang_2012}.
However, the addition of perpendicular diffusion will be the subject of further studies.

We consider pitch-angle scattering mean free paths in the range $\lambda = 0.1$ to $1.0$ au and simulate particle propagation for three days with injection occurring over the first two days from the shock-like source. We consider a monoenergetic population of 5 MeV protons. In each simulation we inject $N_p = 10^6$ particles. An isotropic velocity distribution at injection is assumed. While in our previous modelling work particles were injected into the simulation at $t = 0$ and close to the Sun (instantaneous injection), here we use the model of shock-like injection described in section \ref{sec:Inj} to start particles at a distance $r$ and $t = r / v_{sh}$.

\section{Geometry of observers}\label{sec:Geometry}
\begin{figure*}
    \centering
    \includegraphics[width = 1.0\linewidth, keepaspectratio = true]{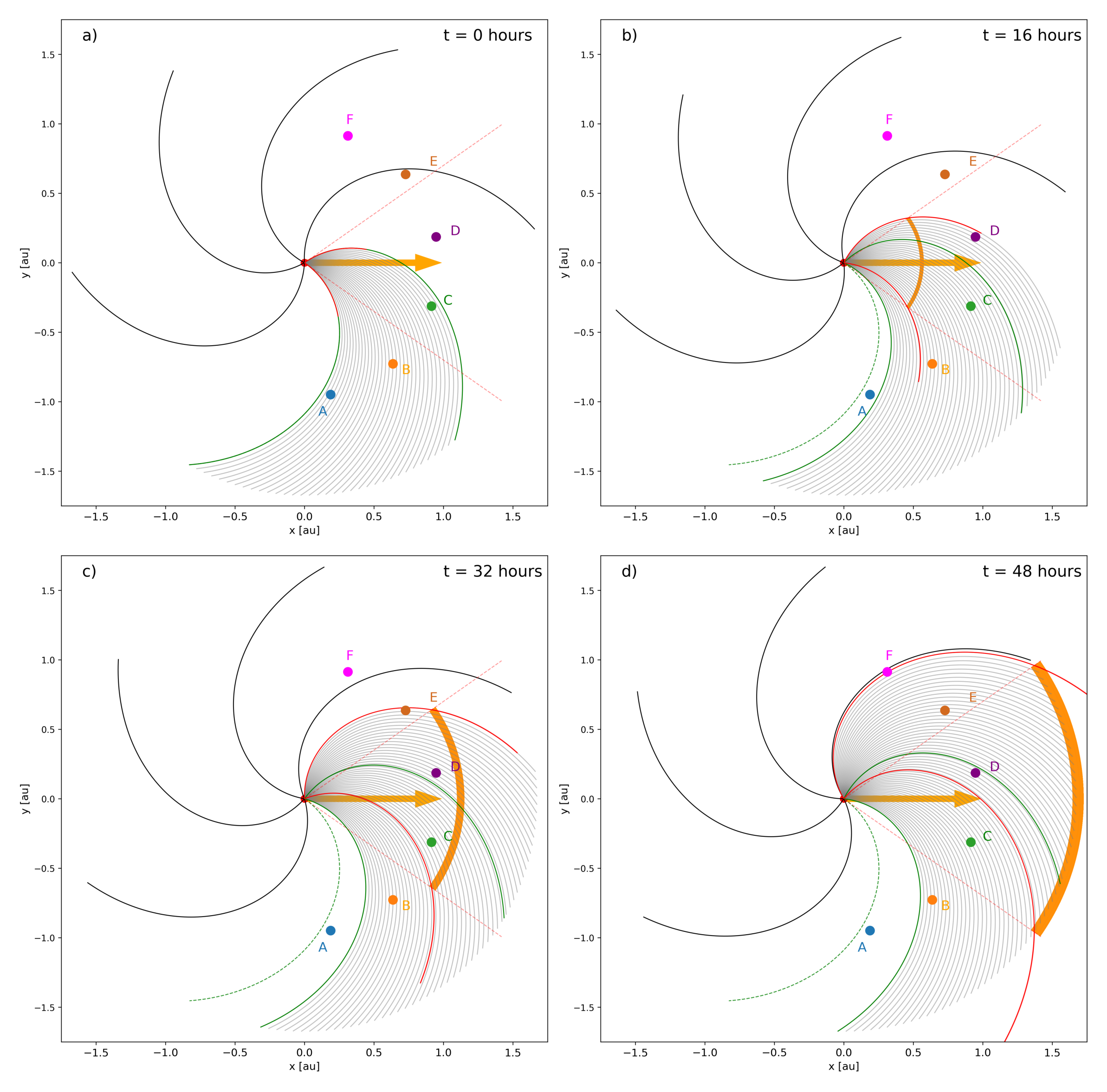}
    \caption{Diagrams showing the shock position and observer geometry at t = 0 (top left), 16 (top right), 32 (bottom left) and 48 (bottom right) hours projected onto the solar equatorial plane. Here x and y are heliocentric cartesian coordinates in the heliographic equatorial plane. The shock's projection onto the plane is displayed here as the orange shaded segments. Observers A-F are denoted by the coloured circles, their exact positions are displayed in Table \ref{obs_table}. The red radial dashed lines delimit the bounds of the shock, and the solid red curved lines show the IMF lines that are currently connected to the flanks of the shock front. The solid green curved lines show the bounds of the shock-like injection at the initial time (or equally an instantaneous injection at the Sun of the same angular width). The dashed green line shows the original position of the left most solid green line at the initial time. The grey IMF lines represent the range of IMF lines that have had particles injected onto them (the particle-filled flux tubes). }
    \label{movie_plots}
\end{figure*}

We derive observables at six observers (labelled A-F) located at 1 au, which can be seen in Figure \ref{movie_plots}, describing the observer-shock geometry at four snapshots in time. In Figure \ref{movie_plots} the orange arrow gives the longitude of the shock nose (coincident with the Active Region (AR) longitude on the Sun), the thin grey curved lines show the range of flux tubes that have been filled with particles by the shock up to time $t$. The solid green lines delimit the longitudinal range of IMF lines that are connected to the shock at the initial time (or an instantaneous injection at the Sun with the same angular width), and the red solid curved lines show the range of IMF lines connected to the shock front at the current time.

The AR is located at ($0^\circ , 15^\circ$) longitude and latitude respectively. As specified in Table \ref{obs_table}, observers \textit{A}, \textit{B} and \textit{C} observe the AR as western, while observers \textit{D}, \textit{E} and \textit{F} observe it as eastern. Observers \textit{C} and \textit{D} are in the path of the shock and will observe it as it passes them, while the other observers will not experience the shock passage in-situ. All observers are located at $\delta = 15^\circ$ latitude to enable connection to the nose of the shock. We have examined the intensity profiles for observers at the same longitudes and at latitude $\delta = 0^\circ$ and there are only very minor differences compared to the plots for $\delta = 15^\circ$, shown in section \ref{sec:int_profs}.

We define the longitudinal separation between the AR and the observer footpoint, $\Delta \phi$ as, 
\begin{equation}
    \Delta \phi = \phi_{AR} - \phi_{ftpt}
    \label{dphi_eqn}
\end{equation}
where $\phi_{AR}$ is the longitudinal position of the source AR on the Sun and $\phi_{ftpt}$ is the longitude of the footpoint of the IMF line connected to the observer. Values of $\Delta \phi$ for our observers are given in Table \ref{obs_table}.

An observer's connection to the shock evolves over time. The observer's cobpoint, where the observer's IMF line meets the shock, moves eastward along the shock front as the shock propagates outwards \citep[e.g.][]{heras_1994,Kal&Wib_1997}.

\begin{table}[h!]
    \centering
    \caption{Observers A-F shown in Figure \ref{movie_plots}. Columns are from left to right: Observer label, location of AR source of the event with respect to observer, and $\Delta \phi$ (see Equation \ref{dphi_eqn}). All observers are located at latitude $\delta = 15^\circ$.}
    \begin{tabular}{ c c c }
        \hline
        {Observer}  & {AR location} & {$\Delta \phi$ [$^\circ$]} \\
        \hline 
         
        A  & W79 & 30 \\
        
        B & W49 & 0 \\
        
        C  & W19 & -30 \\
        
        D  & E11 & -60 \\
        
        E  & E41 & -90 \\
        
        F  & E71 & -120 \\
        \hline

    \end{tabular}
    
    \label{obs_table}
\end{table}

\section{Intensity time profiles and anisotropies}\label{sec:int_profs}

From the output of our test particle simulations we have calculated observables such as intensity profiles and anisotropies for our observers under a number of conditions. Particle counts at each observer are collected over a $10^\circ \times 10^\circ$ tile in longitude and latitude.
Particle anisotropy $A$, is calculated at each observer using the following equation \citep[e.g.][]{Kal&Wib_1997},

\begin{equation}
    A = \frac{3 \int_{-1}^{1} f(\mu) \: \mu \: d\mu}{\int_{-1}^1 f(\mu) \:d\mu}  
    \label{anisotropy_eqn}
\end{equation}
where $f(\mu)$ is the pitch angle distribution and $\mu$ is the pitch angle cosine. We note that the sign of the anisotropy is dependent on the magnetic polarity: in the following we used a unipolar antisunward magnetic field in which case positive anisotropy represents antisunward propagating particles. For this polarity according to Equation \ref{anisotropy_eqn} an anisotropy of 3 corresponds to a fully beamed population travelling along the IMF antisunward.  

\subsection{Effect of scattering mean free path on intensity and anisotropy profiles}
\begin{figure*}
    \centering
    \includegraphics[width = 0.7\linewidth, keepaspectratio = true]{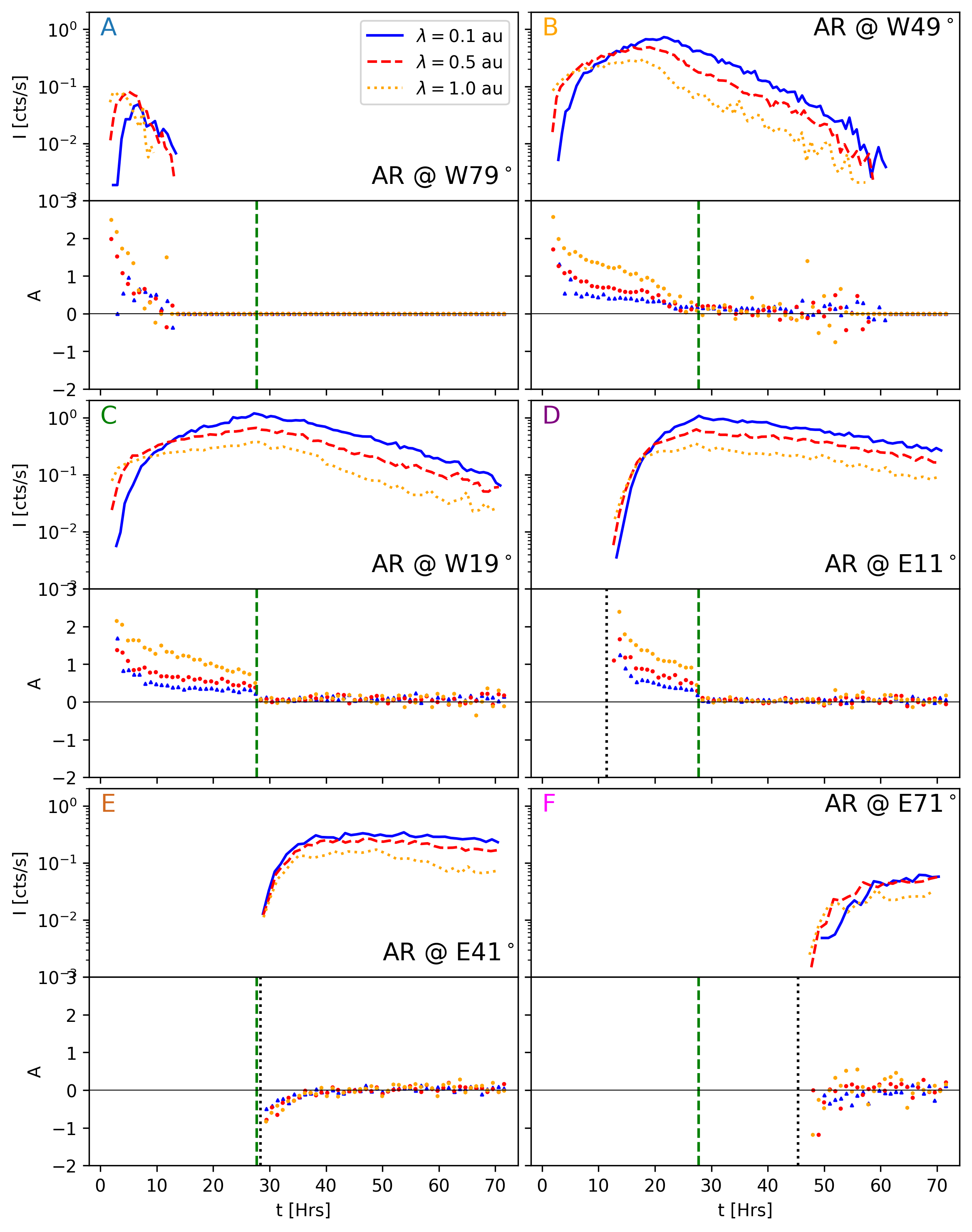}
    \caption{Intensity profiles of a monoenergetic population of 5 MeV protons for observers \textit{A}-\textit{F} at 1.0 au for a range of scattering conditions from $\lambda = 0.1$ au to $\lambda = 1.0$ au, for uniform injection in $r, \theta$ and $\phi$. The green dashed line indicates the time at which the shock reaches the observer radial distance, and the vertical dotted line shows the time when the observer establishes connection to the shock, for observers not connected at the initial time.}
    \label{mfp-1au-comp}
\end{figure*}

We begin by analysing intensities and anisotropies at observers \textit{A} - \textit{F} under a variety of scattering conditions. Here we consider uniform injection in $r,\theta,\phi$.
In Figure \ref{mfp-1au-comp} we display the intensity and anisotropy profiles for the six observers, for simulations with mean free path spanning an order of magnitude from $\lambda = 0.1$ to $1.0$ au.  

In Figure \ref{mfp-1au-comp} intensity profiles for different $\lambda$-values at each observer appear remarkably similar to each other. In contrast to what has been derived from traditional 1D focussed transport models \citep[e.g.][]{Bieber_1994}, we do not see a significant change in the decay time constants with $\lambda$. SEPs in simulations with larger $\lambda$ stream out more quickly. As expected for larger $\lambda$ we find smaller peak intensities and larger anisotropies, due to fewer scattering events. 

The broad features of the intensity profiles at the six observers can be understood in terms of the geometry of the observers relative to the shock. As seen in Figure \ref{movie_plots}a, observers \textit{A}, \textit{B} and \textit{C} are connected to the shock at the initial time. As the shock propagates outwards their cobpoints fall off the eastern edge of the shock, losing connection to the particle source (see Figure \ref{movie_plots}d for observer \textit{A}). Over time the shock connects to longitudes further west and allows observers \textit{D}, \textit{E} and \textit{F} to become connected to the shock-source \citep[e.g.][]{heras_1994}. This enables observers \textit{D}, \textit{E} and \textit{F} that are not initial magnetically connected to the shock to observe SEPs once connection is established. The intensity profiles in Figure \ref{mfp-1au-comp} reflect the different timings of the observer-shock connection, with the onset of the event being delayed at observers \textit{D}-\textit{F}. 

In Figure \ref{mfp-1au-comp} the bottom panels display the anisotropy profiles observed at each of the six observers. 
The anisotropies clearly indicate the times of observer-shock connection. In Figure \ref{mfp-1au-comp} the vertical green line shows the time of shock arrival at the observer's radial distance and the vertical black dotted line indicates the nominal time of observer-shock connection, for the observers not connected at the initial time. For observers \textit{C} and \textit{D} sustained long-duration anisotropies are observed until the time of shock passage, in agreement with previous studies \citep[e.g.][]{Kal&Wib_1997,Kallenrode_2001}. Observers \textit{E} and \textit{F} become connected to the shock when it is located beyond the observer radial position, and so see negative anisotropies due to the sunward propagating particles from the shock.


Observers \textit{C} and \textit{D} see the peak intensity at the time of shock passage. Once the shock propagates past these observers they are receiving sunward propagating particles injected at the shock and particles that were previously injected into the flux tube and have experienced scattering. As the shock propagates beyond the observer the intensity of the former component diminishes with time as particles must overcome the magnetic mirror effect to reach the observer \citep[e.g.][]{Kle2018,Hutch_2022}. As a result after shock passage the anisotropies drop to zero, similar to the results of \cite{Kal&Wib_1997}.

The peak intensities for observers \textit{A} and \textit{B} coincide with their loss of connection to the shock front due to corotation and shock radial motion (i.e. the observer's cobpoint falls off the eastern edge of the shock front). These geometric effects are highly dependent on parameters such as the shock longitudinal width, $w_{sh,\phi}$, the radial speed of the shock, and the observer longitudinal position relative to the edge of the shock front. \cite{hutch_2022-in_prep} compared observables for the cases with and without corotation included, and demonstrated that corotation plays a major role in the decay phase of the event.

\subsection{Time-extended vs instantaneous injection}

\begin{figure}
    \centering
    \includegraphics[width = 1.0\linewidth, keepaspectratio = true]{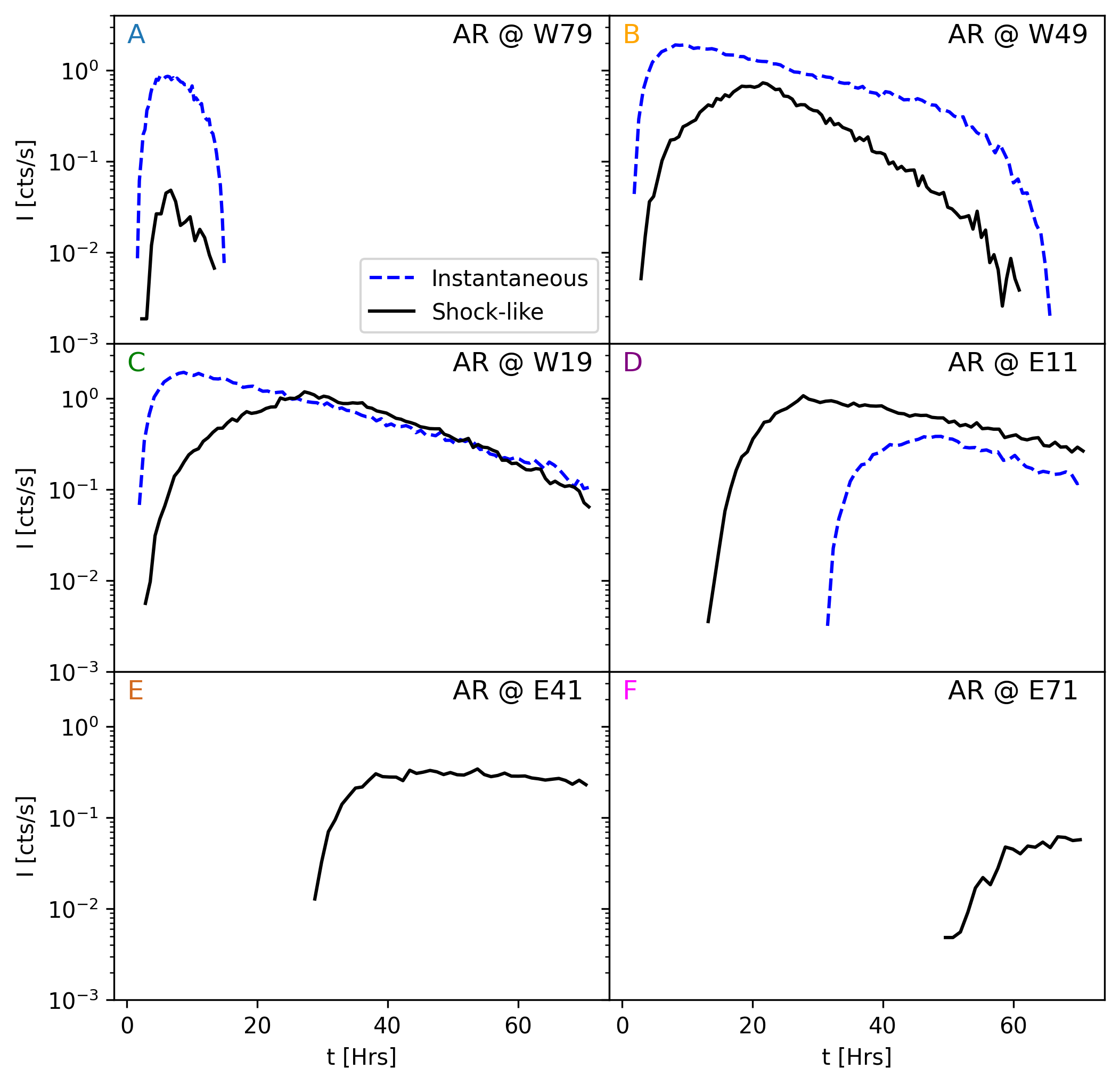}
    \caption{Intensity profiles for observers \textit{A}-\textit{F} considering an instantaneous injection (blue dashed lines) and an extended uniform injection (black solid lines) with respect to $r,\: \theta$ and $\phi$. The injection region is $70^\circ$ in both cases. The simulations use $\lambda = 0.1$ au.}
    \label{inst-unif-plt}
\end{figure}

In previous work with our test-particle model we only considered an instantaneous injection close to the Sun. We now compare observables for the cases of instantaneous injection and a radially uniform time-extended shock-like injection. We consider an instantaneous injection that has the same angular width as the extended injection ($70^\circ \times 70^\circ$). The intensity profiles for the two cases can be seen in Figure \ref{inst-unif-plt} for simulations considering $\lambda = 0.1$ au, where the solid black line is the  radially uniform shock-like injection and the dashed blue line is the instantaneous injection. The same number of particles were injected into each simulation resulting in a significantly larger particle density close to the Sun for the instantaneous injection. This leads to systematically larger intensities for well-connected observers for the instantaneous injection (observers \textit{A}-\textit{C}). From the first panel (Observer \textit{A}) it can be seen that both the time-extended and instantaneous injection lead to the same event duration ($\sim 12$ hours). This occurs as the corotation sweeps the particle-filled flux tubes westward and away from the observer. Similarly, for observer \textit{B} the intensity profile is cut short by the corotation. The corotation enables observer \textit{D} to see a signal from the instantaneous injection as the particle-filled flux tubes corotate to this observer. In the shock-like simulations observer \textit{D} detects SEPs from the CME-driven shock approximately 10 hours earlier compared to the instantaneous injection, showing the clear timescale differences between the two cases. Observers \textit{E} and \textit{F} do not receive particles from the instantaneous injection over the timescales of our simulations as the corotation takes longer than 72 hours to rotate the particle-filled flux tubes to these observers. From Figure \ref{inst-unif-plt} it is clear that a temporally extended injection does not necessarily mean a longer duration SEP event, especially for western events.

\subsection{Dependence on radial injection function}

\begin{figure}
    \centering
    \includegraphics[width = 1.0\linewidth, keepaspectratio = true]{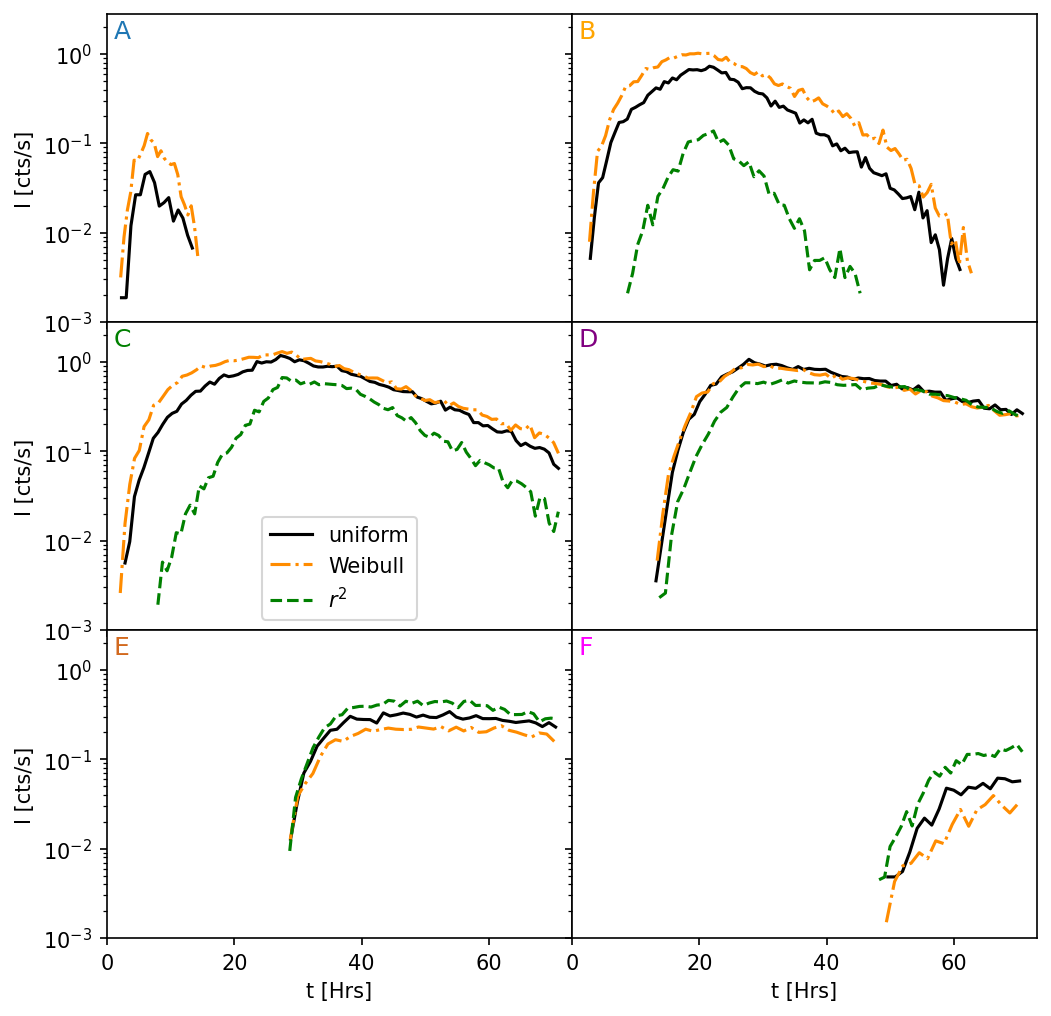}
    \caption{Intensity profiles for the three radial injection functions of Figure \ref{Inj_rad_plt} at each of the six observers in Table \ref{obs_table} with scattering conditions described by $\lambda = 0.1$ au. Injection is uniform in $\theta, \: \phi$.}
    \label{rad_inj_int_prof}
\end{figure}

In Figure \ref{rad_inj_int_prof} we consider three radial injection functions (uniform, Weibull and $r^2$, see Figure \ref{Inj_rad_plt}) and determine the intensity profiles at each of the six observers.
The intensity profiles are surprisingly similar considering the very different radial injections. Comparing the uniform and Weibull function injections it can be seen that there is little difference in the intensity profiles, only showing minor differences in the rise phase and peak intensities for western events (\textit{A}, \textit{B}, \textit{C}) due to the increased particle numbers injected close to the Sun for the Weibull injection. The very similar intensity profiles imply that the number of particles injected per unit area of the shock front, $Q(r)$ has a larger effect on intensity profiles than $R(r)$. The very low particle numbers for the $r^2$ injection close to the Sun results in no observable signal at observer \textit{A} for this injection. This is also the reason for the smaller intensities observed at observers \textit{B} and \textit{C}.

For observer \textit{D} all three injections are very similar, with the $r^2$ injection having slightly lower intensities during the rise phase. For more eastern events (observers \textit{E} and \textit{F}) the $r^2$ injection shows the largest intensities due to these observers connecting to the shock at larger radial distances where larger numbers of particles are injected. 
It is clear from these plots that geometric effects such as the time of observer-shock connection and the corotation/disconnection of particle-filled flux tubes toward/away from the observer has a much more significant effect on the intensity profile than the radial injection function.

\begin{figure}
    \centering
    \includegraphics[width = 1.0\linewidth, keepaspectratio = true]{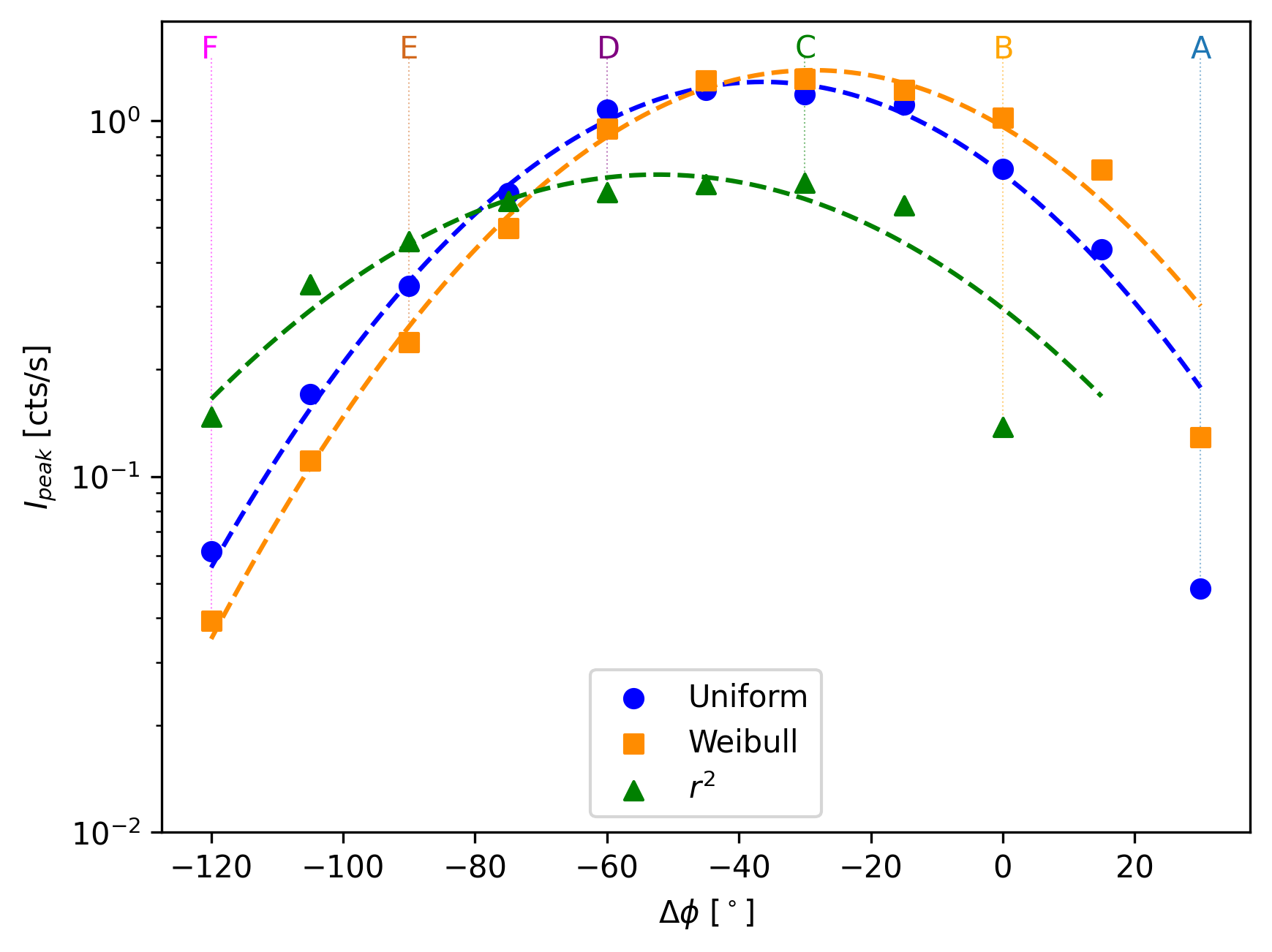}
    \caption{Peak intensity versus $\Delta \phi$ for the three radial injection functions.}
    \label{Ipeak}
\end{figure}

The peak intensity, $I_{peak}$ is plotted in Figure \ref{Ipeak} versus $\Delta \phi$ (defined in Equation \ref{dphi_eqn}) for the three shock-like radial injection functions. Each set of points is fit with a Gaussian of the form $I = I_0 \exp(-(\phi - \phi_0)^2/2\sigma^2)$. Table \ref{table:gauss_table} shows the values of $\phi_0$, $\sigma$ and $I_0$ for the fits shown in Figure \ref{Ipeak}. The largest intensities are obtained for the Weibull $R(r)$, which injects most particles close to the Sun, while the $r^2$ injection function results in much lower $I_{peak}$ values. The standard deviations of the uniform and Weibull function injections are similar, while the $r^2$ injection has a larger standard deviation. The peak intensities at observers \textit{B} and \textit{C} are lower for the $r^2$ injection compared to the other injection functions as the number of particles injected close to the Sun is smaller. However, at observers \textit{D} \textit{E} and \textit{F} larger intensities are seen because the shock continues to inject particles late into the event. Figure \ref{Ipeak} also shows that the broadness of the Gaussian is more closely related to the number of particles injected per unit area of the shock front, $Q(r)$, rather than $R(r)$. The centre of the Gaussian, $\phi_0$, is shifted towards more negative values as one goes from Weibull to uniform to $r^2$ $R(r)$.  

\begin{table}[h]
    \centering
    \caption{Parameter values of the Gaussians fitted to the peak intensities versus $\Delta \phi$ plots in Figures \ref{Ipeak} and \ref{fig:sigma_gauss}.}
    \begin{tabular}{c c c c c}
        \hline
        {$R(r)$}  & {$\Phi(\phi)$ $\Theta(\theta)$} & {$I_0$} & {$\phi_0$ [$^\circ$]} & {$\sigma$ [$^\circ$]}\\
        \hline 
         Uniform & Uniform & 1.28 & -36.4 & 33.4 \\
         Weibull & Uniform &  1.39 & -28.8 & 33.6 \\
         $r^2$ & Uniform & 0.71 & -53.5 & 37.6 \\
         Uniform & Gaussian & 2.55 & -38.1 & 25.6 \\
        \hline
    \end{tabular}
    
    \label{table:gauss_table}
\end{table}

\subsection{Dependence on varying injection efficiency across the shock}
\begin{figure}
    \centering
    \includegraphics[width = 1.0\linewidth, keepaspectratio = true]{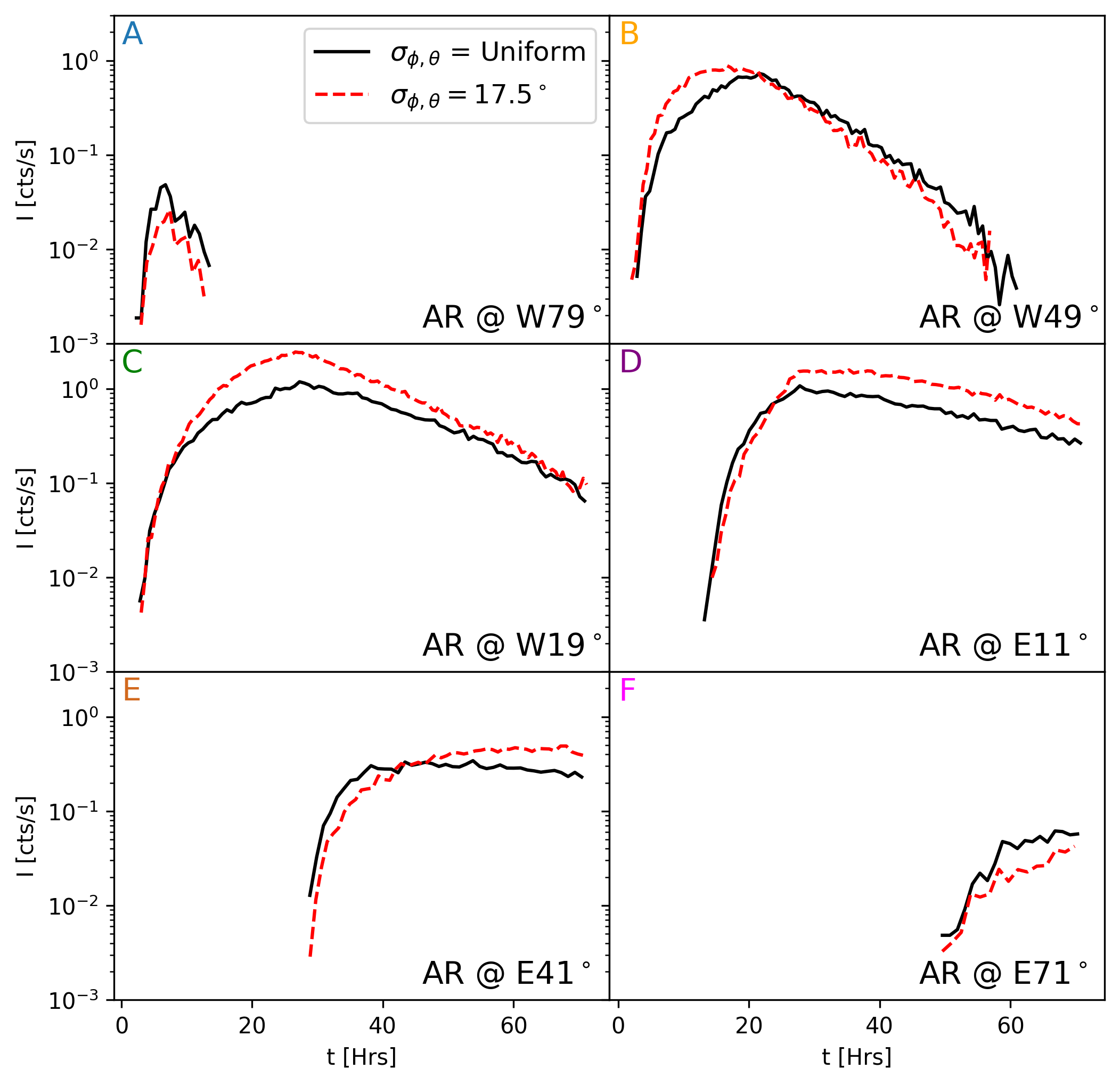}
    \caption{Intensity profiles for observers \textit{A}-\textit{F} at 1 au considering uniform and Gaussian longitudinal and latitudinal injection functions, with $\lambda = 0.1$ au. Injection is uniform in $r$.}
    \label{unif-gauss-int-profs}
\end{figure}

In Figure \ref{unif-gauss-int-profs} intensity profiles are shown when considering injection efficiencies across the shock in longitude and  colatitude ($\Phi(\phi), \: \Theta(\theta)$) that are a) uniform (black solid line) and b) Gaussian (red dashed line) with  $\sigma = 17.5^\circ$ centred on the shock nose (see Figure \ref{long_func}).
It is apparent that changing the injection efficiency across the shock produces only small changes in the intensity profiles. For the Gaussian injection function at times when the observer's cobpoint lies near the shock nose (observers \textit{B} and \textit{C}) there is a faster increase compared to the uniform case during the rise phase, and a larger peak intensity. Observers \textit{A} and \textit{F}, which experience connection to the flanks of the shock, have smaller intensities compared to the uniform injection case, but the differences are small. Changing the injection profiles across the shock does not have a big influence on the overall features of the intensity profiles. 

\begin{figure}
    \centering
    \includegraphics[width = 1.0\linewidth, keepaspectratio = true]{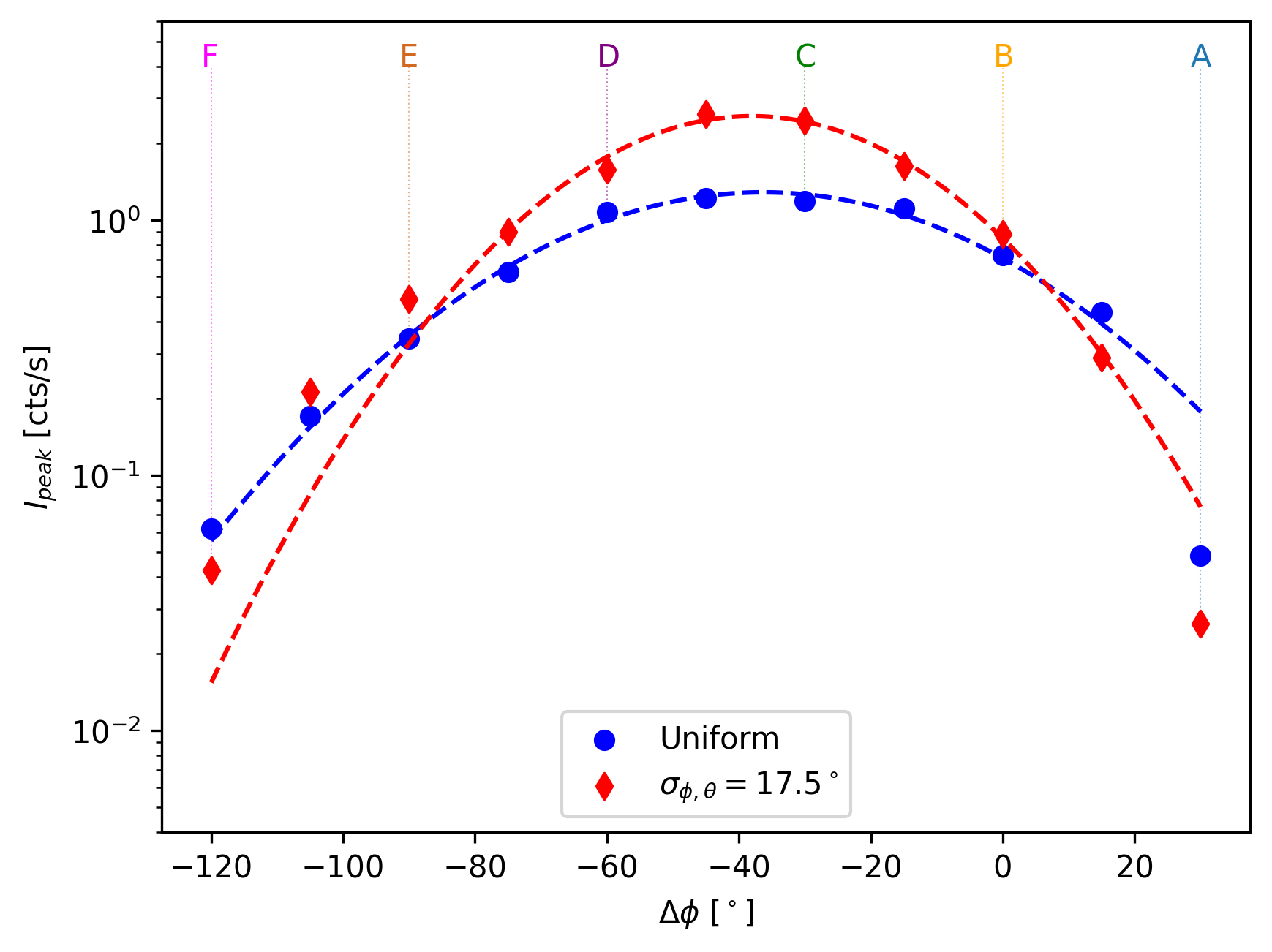}
    \caption{Peak intensity versus $\Delta \phi$ for shocks with uniform (blue circles) or Gaussian (red diamonds, $\sigma = 17.5^\circ$) longitudinal and latitudinal injection efficiency. The shocks have an angular width of $70^\circ$. Both sets of points are fitted with Gaussians (dashed lines).}
    \label{fig:sigma_gauss}
\end{figure}

In Figure \ref{fig:sigma_gauss} we plot the peak intensity versus $\Delta \phi$ for uniform and Gaussian injection efficiency across the shock. Both sets of points are fitted with a Gaussian and the corresponding fit parameters are given in Table \ref{table:gauss_table}. As expected, having a Gaussian injection reduces the standard deviation of the fit due to the more spatially confined injected population.

\subsection{Role of shock width}
\begin{figure}
    \centering
    \includegraphics[width = 1.0\linewidth, keepaspectratio = true]{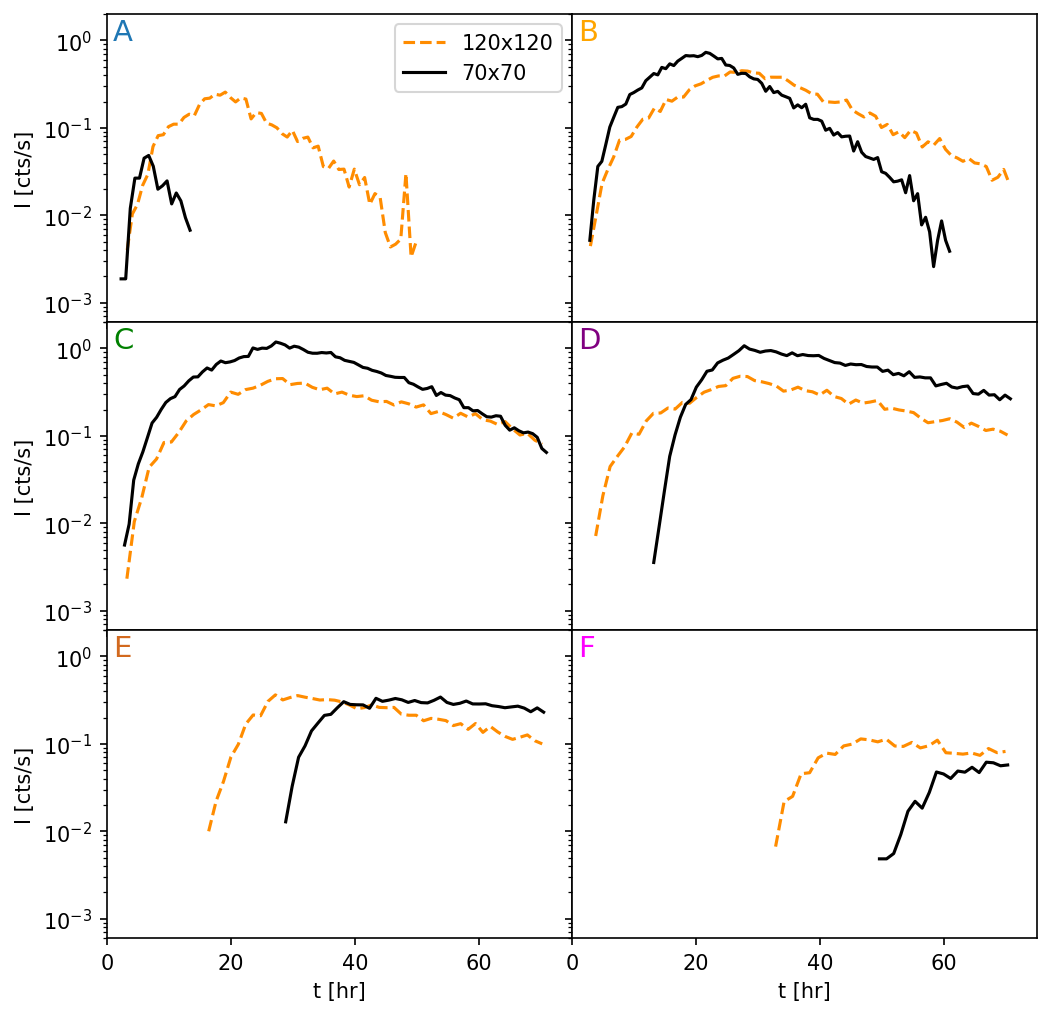}
    \caption{Intensity profiles at the six observers considering a shock $70^\circ$ (solid black line) and $120^\circ$ (dashed orange line) in both longitude and latitude.}
    \label{width_plt}
\end{figure}

We have considered shocks of different angular widths. In Figure \ref{width_plt} we compare intensity profiles at the six observers for shocks with $w_{sh,\phi} = w_{sh,\theta} = 70^\circ$ (black line) and $120^\circ$ (orange dashed line). 

One effect of a wider shock is that an observer can remain magnetically connected to the shock front for a longer period of time. This factor changes the rise times for western events because the peak position in the intensity profiles is determined by the loss of connection to the shock front. This can be seen in Figure \ref{width_plt} for observers \textit{A} and \textit{B} where the rise times are extended and peak intensities occur later. Observations at \textit{C} are similar in the two cases as the peak intensity occurs as the shock passes directly over the observer. We note that the exact values of the intensities are not directly comparable in these simulations as the same number of particles are injected over a larger volume for the wider shock (i.e. different $Q(r)$).
For a wider shock at observers that see the event as eastern, onsets take place earlier as the observer-shock connection is established more quickly, when the shock is closer to the Sun, as can be seen in Figure \ref{width_plt} for observers \textit{D}, \textit{E} and \textit{F}.

\subsection{Observers at 0.3 au}
Here we use our simulations to obtain observables close to the Sun and compare them with 1 au observables. This is important now that \textit{Parker Solar Probe} and \textit{Solar Orbiter} are obtaining in-situ data in the inner heliosphere.   

\begin{figure*}
    \centering
    \includegraphics[width = 1.0\linewidth, keepaspectratio = true]{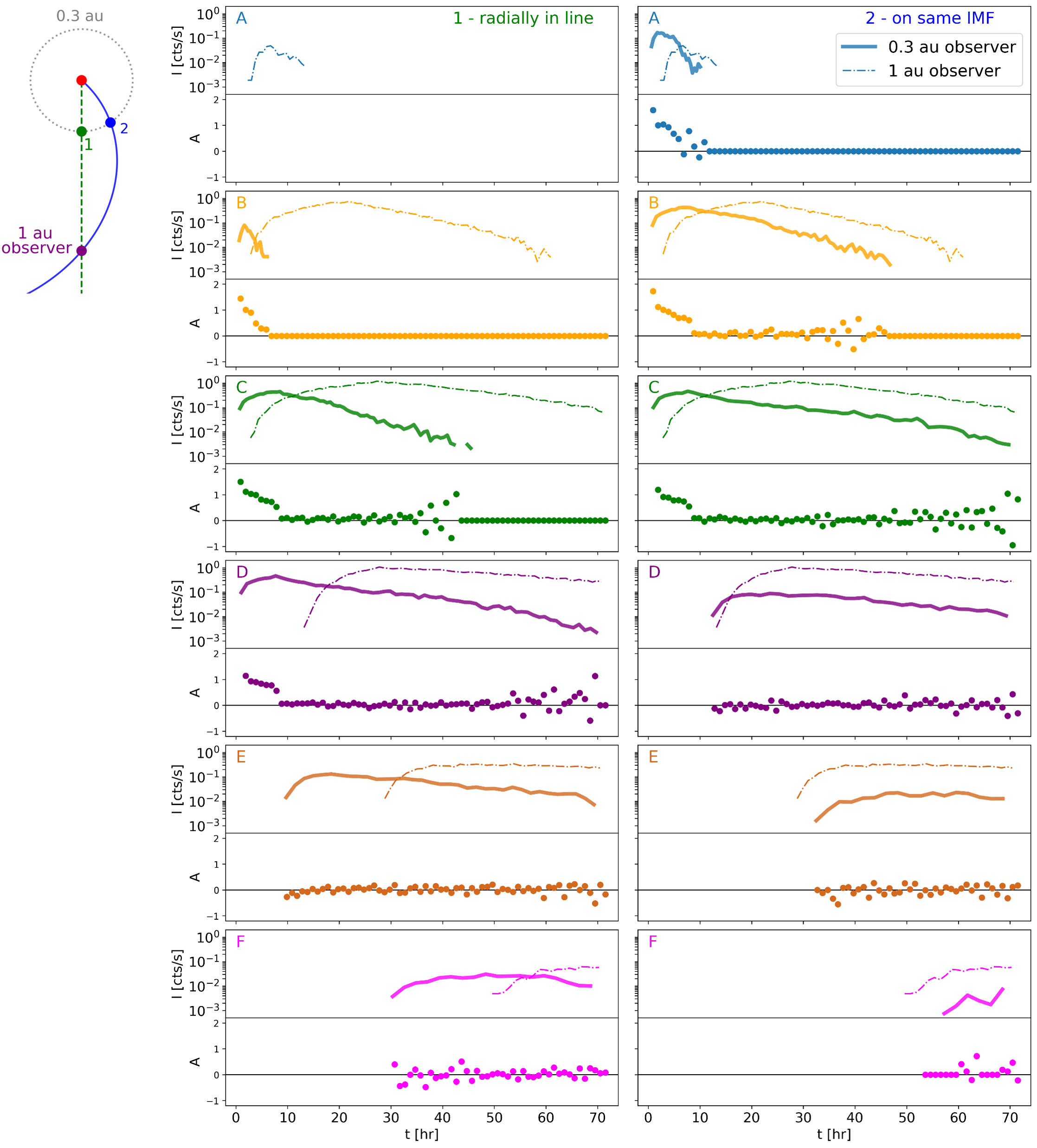}
    \caption{Left: Schematic of observer geometry, with observer 1 located at 0.3 au, radially aligned with the 1 au observers, and 2 located at 0.3 au along the same IMF line as the 1 au observer. Right: Intensity and anisotropy profiles for observers 1 and 2 (solid curves) and 1.0 au observer (dash-dotted curves). }
    \label{12-panel-0_3-1au}
\end{figure*}

Having fixed an observer at 1 au, in Figure \ref{12-panel-0_3-1au} intensity and anisotropy profiles are plotted for two 0.3 au observers: the first radially in line with the 1 au observer (left column) and the second on the same IMF line (right column). Profiles at the 1 au observer are indicated by the dash-dotted lines for comparison. A uniform $R(r)$ is used.

Compared to the 1 au profiles, the $0.3$ au profiles are characterised by: faster rise times, due to faster shock passage times at 0.3 au, and larger anisotropies (1 au anisotropies not shown) as there is reduced isotropisation due to fewer pitch-angle scattering events.

Generally the profiles at 0.3 and 1.0 au appear more similar for the case of observers on the same IMF line (right column), than for the radially in line observers (left column). In the former case both observers establish connection to the shock at similar times, with some difference due to propagation times. For the case of radially in line 0.3 and 1.0 observers (left column) there are stronger differences between the profiles compared to the previous case. For observers \textit{D}, \textit{E} and \textit{F} the onset at 0.3 au is more than $10$ hours earlier than at 1 au. This is due to the fact that the radially aligned 0.3 au observers have footpoints located eastward of the 1 au observer, meaning that for eastern events they will connect to the shock earlier. The decay time constants of the event appear significantly different at 0.3 and 1.0 au. 

Considering peak intensities for the same IMF line case, for observers \textit{B} - \textit{F} the peak intensity at 1.0 au is larger than at 0.3 au. This is because particle injection before shock arrival is more extended for the 1 au observer. After shock passage, although particles can propagate sunward to the observer, this becomes difficult due to magnetic mirroring \citep{Hutch_2022}. For the observer \textit{A} panel the 0.3 au intensity is larger because both 0.3 and 1 au observers lose connection early in the event and scattering and propagation delay result in lower peak intensity at 1 au. The intensity profiles observed at 0.3 au have a weak dependence on $\lambda$ (not shown).
Eastern events show much slower decay phases compared to western events, similar to the 1.0 au observers. 

We note that when close to the Sun \textit{Parker Solar Probe} and \textit{Solar Orbiter} are moving in longitude at high speed so that the intensity profiles shown in Figure \ref{12-panel-0_3-1au}, where the $0.3$ au observer is stationary, will not correspond exactly to those measured by these spacecraft.  In some phases of the mission \textit{Solar Orbiter} will be corotating with the Sun.

We have analysed the intensity profiles (not shown) at $0.3$ au for the three radial injection profiles considered earlier (see Figure \ref{Inj_rad_plt}) and they display a behaviour similar to that in Figure \ref{rad_inj_int_prof}, i.e. not a strong dependence on $R(r)$.

\section{Discussion and conclusions}\label{sec:disc}


In this paper we have presented the first 3D test particle simulations of SEPs with a temporally extended shock-like particle injection. Previously, in our modelling we had only considered an instantaneous injection near the Sun. By deriving the intensity and anisotropy profiles for observers at 0.3 and 1.0 au, we have reached the following main conclusions:

\begin{enumerate}
    \item The main difference between an instantaneous and time-extended injection is that in the former case the spatial extent of the accelerated particle population is smaller (Figure \ref{inst-unif-plt}). For initially well-connected observers (\textit{A}-\textit{C}) the duration of the SEP event is not significantly shorter for an instantaneous injection compared to an extended one.

    
    
    
    \item The radial profile of injection (radial injection function, $R(r)$, Figure \ref{Inj_rad_plt}) plays a surprisingly small role in determining the intensity profiles at 1 au (Figure \ref{rad_inj_int_prof}). However, $R(r)$ has a strong effect on the heliolongitudinal distribution of peak intensity (Figure \ref{Ipeak}) with injections that continue over larger radial distances leading to more negative $\phi_0$ values and larger standard deviations. 
    
    \item Varying the injection efficiency across the shock (longitudinal and latitudinal injection functions, $\Phi(\phi)$, $\Theta(\theta)$, Figure \ref{long_func}) also plays a  minor role in shaping intensity profiles (Figure \ref{unif-gauss-int-profs}).
    
    \item In most cases simulations show large persistent anisotropies prior to shock passage and they decay sharply at shock arrival, becoming very close to zero following shock passage (Figure \ref{mfp-1au-comp}, observers \textit{A}-\textit{D}). For observers that see the event far in the east the first arriving particles are propagating sunward once connection to the shock is established.
    
    \item Larger shock widths lead to longer duration SEP events because the observers remain connected to the shock for a longer time.

    \item Intensity profiles at 0.3 au are similar to those at 1.0 au for two observers on the same IMF line, but show faster rise times, and larger anisotropies.

\end{enumerate}
Our simulations show that the link between duration of injection and duration of the SEP event is very weak, unlike what is commonly assumed. Also from our simulations it is not clear that differences in the acceleration efficiencies at the flanks of the shock would leave a signature in the observed intensity profiles, as is often postulated \citep[e.g.][]{Tylka_2006}.
Spatial/geometric factors such as the establishment/loss of observer-shock connection and the corotation of the particle-filled flux tubes toward/away from the observer are the dominant factors in determining the shapes and properties of SEP intensity profiles. The role of corotation is fully illustrated in a companion paper \citep{hutch_2022-in_prep}.
Intensity profiles show little dependence on mean free path, $\lambda$. In particular the decay phase constant is weakly dependent on $\lambda$, unlike what is derived from 1D focussed transport models \citep[e.g.][]{Bieber_1994}.

A number of studies of SEP observations have derived plots of peak intensity versus the separation $\Delta \phi$ between source AR and observer footpoint \cite[e.g.][]{wied_2013,Lario_2013,cohen2014,Richardson2014}. Our work shows that these plots are sensitive to the radial injection profile and longitudinal/latitudinal injection efficiency. When a lot of acceleration takes place in interplanetary space, the centre $\phi_0$ of the Gaussian fit to the $I_{peak}$ versus $\Delta \phi$ plot is shifted towards more negative values compared to cases where most of the acceleration takes place close to the Sun. For the shock width considered in our study, which is $70^\circ$, standard deviations between 26$^\circ$ and 38$^\circ$ are found. 

In this paper we presented results for a monoenergetic proton population injected with energy of 5 MeV, which is a representative SEP energy.  The difficulty in considering multiple energies lies in the need to specify how the radial injection function, $R(r)$, varies with energy. In addition because of particle deceleration, the final particle energy in our simulations is smaller than 5 MeV for some particles. In constructing the intensity profiles in this paper we have chosen to include all particles $>$1 MeV. In actual SEP events a spectrum of energies would actually be injected and particles of higher energy would decelerate into the 5 MeV range. Limiting the range of energies in the plot to a smaller energy range near 5 MeV produces some modifications in the profiles, but does not change the qualitative trends we have found.
For particle energies much higher than 5 MeV, injections are thought to take place only close to the Sun, limiting the range of longitudes of the shock-source. For these higher energy particles, drift effects may be important in determining the range of accessible longitudes during an SEP event \citep{dalla_2013}.

Our model did not include magnetic field line meandering \citep{Laitinen_2016}, which would lead to significant motion of the particles perpendicular to the mean magnetic field. This effect could potentially explain some features of our modelled intensity profiles that do not agree with SEP observations: for eastern events our simulations do not show the slow rise phase found in observations \citep[e.g.][]{cane_1988,Kahler_2016}, displaying instead very delayed onsets and relatively fast rises. Inclusion of perpendicular transport may produce the observed long rise times as long as the process is slow. It might be possible to determine a limit to the strength of perpendicular transport by modelling the slow rise times during eastern events. 
Considering the plots of $I_{peak}$ versus $\Delta \phi$ (Figures \ref{Ipeak} and \ref{fig:sigma_gauss}) perpendicular diffusion would increase the peak intensity at the less well-connected observers, resulting in a larger standard deviation of the Gaussian fits. It is expected that with the inclusion of perpendicular diffusion intensity profiles at widely separated locations will become more similar to each other. 
We hope to include the effects of turbulence-induced perpendicular transport in future work. 

The anisotropy characteristics shown in Figure \ref{mfp-1au-comp}, namely the large persistent anisotropies and the sunward anisotropies for far eastern events are not routinely detected in SEP events, to our knowledge. It is possible that this may be due to perpendicular transport effects.

While our model of shock-like injection has allowed us to derive the qualitative patterns described above, it contains several simplifications that will need to be improved upon in future work. Our simulation does not model shock acceleration, nor the interaction of energetic particles with the shock.
We have not considered how the shock decelerates with time: this would affect the extent of the event since IMF field lines towards the west would only be reached at later times.  

The simulations in the present study do not include the Heliospheric Current Sheet (HCS), which has been shown to have significant effects for SEP propagation when the source region is located close to it \citep{Battarbee_2017, Battarbee_2018,Waterfall_2022}. Individual events, depending on the magnetic configuration, may be significantly influenced by the HCS due to strong HCS drift motions, especially for high energy SEPs \citep{Waterfall_2022}.

Another factor which may influence intensity profiles are complex local magnetic field and solar wind structures, not included in our model at present. Any perturbations to the Parker spiral will affect the times of observer-shock connection and so will affect observable parameters such as onset times and peak times. Some structures like corotating interaction regions \citep{Wijsen_2019} or magnetic clouds \citep{kallenrode_2002} may significantly affect SEP transport affecting intensity profiles.

\bibliographystyle{aa}
\bibliography{methodology}

\begin{thebibliography}{35}
\expandafter\ifx\csname natexlab\endcsname\relax\def\natexlab#1{#1}\fi

\bibitem[{{Baring}(1997)}]{Baring_1997}
{Baring}, M.~G. 1997, in Very High Energy Phenomena in the Universe; Moriond
  Workshop, ed. Y.~{Giraud-Heraud} \& J.~{Tran Thanh van}, 97

\bibitem[{Battarbee {et~al.}(2017)Battarbee, Dalla, \& Marsh}]{Battarbee_2017}
Battarbee, M., Dalla, S., \& Marsh, M.~S. 2017, \apj, 836, 138

\bibitem[{Battarbee {et~al.}(2018)Battarbee, Dalla, \& Marsh}]{Battarbee_2018}
Battarbee, M., Dalla, S., \& Marsh, M.~S. 2018, The Astrophysical Journal, 854,
  23

\bibitem[{{Bieber} {et~al.}(1994){Bieber}, {Matthaeus}, {Smith}, {Wanner},
  {Kallenrode}, \& {Wibberenz}}]{Bieber_1994}
{Bieber}, J.~W., {Matthaeus}, W.~H., {Smith}, C.~W., {et~al.} 1994, \apj, 420,
  294

\bibitem[{Cane {et~al.}(1988)Cane, Reames, \& von Rosenvinge}]{cane_1988}
Cane, H.~V., Reames, D.~V., \& von Rosenvinge, T.~T. 1988, Journal of
  Geophysical Research: Space Physics, 93, 9555

\bibitem[{{Cohen} {et~al.}(2014){Cohen}, {Mason}, {Mewaldt}, \&
  {Wiedenbeck}}]{cohen2014}
{Cohen}, C.~M.~S., {Mason}, G.~M., {Mewaldt}, R.~A., \& {Wiedenbeck}, M.~E.
  2014, \apj, 793, 35

\bibitem[{{Dalla} \& {Browning}(2005)}]{Dalla&Browning_2005}
{Dalla}, S. \& {Browning}, P.~K. 2005, A\&A, 436, 1103

\bibitem[{{Dalla} {et~al.}(2020){Dalla}, {de Nolfo, G. A.}, {Bruno, A.},
  {Giacalone, J.}, {Laitinen, T.}, {Thomas, S.}, {Battarbee, M.}, \& {Marsh, M.
  S.}}]{Dalla_2020}
{Dalla}, S., {de Nolfo, G. A.}, {Bruno, A.}, {et~al.} 2020, A\&A, 639, A105

\bibitem[{Dalla {et~al.}(2013)Dalla, Marsh, Kelly, \& Laitinen}]{dalla_2013}
Dalla, S., Marsh, M., Kelly, J., \& Laitinen, T. 2013, Journal of Geophysical
  Research: Space Physics, 118, 5979

\bibitem[{Desai \& Giacalone(2016)}]{Desai_2016}
Desai, M. \& Giacalone, J. 2016, Living Reviews in Solar Physics, 13

\bibitem[{Gopalswamy {et~al.}(2013)}]{Gop_2013}
Gopalswamy, N. {et~al.} 2013, Adv. Space Res., 51, 1981

\bibitem[{He {et~al.}(2011)He, Qin, \& Zhang}]{He_2011}
He, H.-Q., Qin, G., \& Zhang, M. 2011, \apj, 734, 74

\bibitem[{Heras {et~al.}(1994)Heras, Sanahuja, Sanderson, Marsden, \&
  Wenzel}]{heras_1994}
Heras, A.~M., Sanahuja, B., Sanderson, T.~R., Marsden, R.~G., \& Wenzel, K.~P.
  1994, Journal of Geophysical Research: Space Physics, 99, 43

\bibitem[{{Hutchinson} {et~al.}(2022{\natexlab{a}}){Hutchinson}, {Dalla},
  {Laitinen}, {de Nolfo}, {Bruno}, {Ryan}, \& {Waterfall}}]{Hutch_2022}
{Hutchinson}, A., {Dalla}, S., {Laitinen}, T., {et~al.} 2022{\natexlab{a}},
  A\&A, 658, A23

\bibitem[{{Hutchinson} {et~al.}(2022{\natexlab{b}}){Hutchinson}, {Dalla, S.},
  {Laitinen, T.}, \& {Waterfall, C. O. G.}}]{hutch_2022-in_prep}
{Hutchinson}, A., {Dalla, S.}, {Laitinen, T.}, \& {Waterfall, C. O. G.}
  2022{\natexlab{b}}, A\&A Lett in press (AA/2022/45312), arxiv:2210.15464

\bibitem[{Kahler(2016)}]{Kahler_2016}
Kahler, S.~W. 2016, The Astrophysical Journal, 819, 105

\bibitem[{{Kahler} \& {Ling}(2018)}]{Kahler_2018}
{Kahler}, S.~W. \& {Ling}, A.~G. 2018, \solphys, 293, 30

\bibitem[{Kallenrode(2001)}]{Kallenrode_2001}
Kallenrode, M.-B. 2001, Journal of Geophysical Research: Space Physics, 106,
  24989

\bibitem[{Kallenrode(2002)}]{kallenrode_2002}
Kallenrode, M.-B. 2002, Journal of Atmospheric and Solar-Terrestrial Physics,
  64, 1973

\bibitem[{Kallenrode \& Wibberenz(1997)}]{Kal&Wib_1997}
Kallenrode, M.-B. \& Wibberenz, G. 1997, Journal of Geophysical Research: Space
  Physics, 102, 22311

\bibitem[{{Klein} {et~al.}(2018){Klein}, {Tziotziou}, {Zucca}, {Valtonen},
  {Vilmer}, {Malandraki}, {Hamadache}, {Heber}, \& {Kiener}}]{Kle2018}
{Klein}, K.-L., {Tziotziou}, K., {Zucca}, P., {et~al.} 2018, in Solar Particle
  Radiation Storms Forecasting and Analysis, ed. O.~E. {Malandraki} \& N.~B.
  {Crosby}, Vol. 444, 133--155

\bibitem[{{Laitinen} {et~al.}(2016){Laitinen}, {Kopp}, {Effenberger}, {Dalla},
  \& {Marsh}}]{Laitinen_2016}
{Laitinen}, T., {Kopp}, A., {Effenberger}, F., {Dalla}, S., \& {Marsh}, M.~S.
  2016, A\&A, 591, A18

\bibitem[{Lario {et~al.}(2013)Lario, Aran, G{\'{o}}mez-Herrero, Dresing, Heber,
  Ho, Decker, \& Roelof}]{Lario_2013}
Lario, D., Aran, A., G{\'{o}}mez-Herrero, R., {et~al.} 2013, The Astrophysical
  Journal, 767, 41

\bibitem[{{Lario} {et~al.}(1998){Lario}, {Sanahuja}, \& {Heras}}]{Lario_1998}
{Lario}, D., {Sanahuja}, B., \& {Heras}, A.~M. 1998, \apj, 509, 415

\bibitem[{Marsh {et~al.}(2015)Marsh, Dalla, Dierckxsens, Laitinen, \&
  Crosby}]{Marsh_2015}
Marsh, M.~S., Dalla, S., Dierckxsens, M., Laitinen, T., \& Crosby, N.~B. 2015,
  Space Weather, 13, 386

\bibitem[{Marsh {et~al.}(2013)Marsh, Dalla, Kelly, \& Laitinen}]{Marsh_2013}
Marsh, M.~S., Dalla, S., Kelly, J., \& Laitinen, T. 2013, \apj, 774, 4

\bibitem[{Qin {et~al.}(2013)Qin, Wang, Zhang, \& Dalla}]{Qin_2013}
Qin, G., Wang, Y., Zhang, M., \& Dalla, S. 2013, \apj, 766, 74

\bibitem[{Reames {et~al.}(1997)Reames, Kahler, \& Ng}]{Reames_1997}
Reames, D.~V., Kahler, S.~W., \& Ng, C.~K. 1997, \apj, 491, 414

\bibitem[{Richardson {et~al.}(2014)Richardson, von Rosenvinge, Cane, Christian,
  Cohen, Labrador, Leske, Mewaldt, Wiedenbeck, \& Stone}]{Richardson2014}
Richardson, I.~G., von Rosenvinge, T.~T., Cane, H.~V., {et~al.} 2014, Solar
  Physics, 289, 48

\bibitem[{Tylka \& Lee(2006)}]{Tylka_2006}
Tylka, A.~J. \& Lee, M.~A. 2006, The Astrophysical Journal, 646, 1319

\bibitem[{Wang {et~al.}(2012)Wang, Qin, \& Zhang}]{Wang_2012}
Wang, Y., Qin, G., \& Zhang, M. 2012, The Astrophysical Journal, 752, 37

\bibitem[{Waterfall {et~al.}(2022)Waterfall, Dalla, Laitinen, Hutchinson, \&
  Marsh}]{Waterfall_2022}
Waterfall, C. O.~G., Dalla, S., Laitinen, T., Hutchinson, A., \& Marsh, M.
  2022, The Astrophysical Journal, 934, 82

\bibitem[{{Wiedenbeck} {et~al.}(2013){Wiedenbeck}, {Mason}, {Cohen}, {Nitta},
  {G{\'o}mez-Herrero}, \& {Haggerty}}]{wied_2013}
{Wiedenbeck}, M.~E., {Mason}, G.~M., {Cohen}, C.~M.~S., {et~al.} 2013, \apj,
  762, 54

\bibitem[{{Wijsen} {et~al.}(2019){Wijsen}, {Aran}, {Pomoell}, \&
  {Poedts}}]{Wijsen_2019}
{Wijsen}, N., {Aran}, A., {Pomoell}, J., \& {Poedts}, S. 2019, \aap, 622, A28

\bibitem[{Zank {et~al.}(2006)Zank, Li, Florinski, Hu, Lario, \&
  Smith}]{Zan_2006}
Zank, G.~P., Li, G., Florinski, V., {et~al.} 2006, Journal of Geophysical
  Research: Space Physics, 111

\end{thebibliography}

\begin{acknowledgements}
A. Hutchinson, S. Dalla and T. Laitinen acknowledge support from the UK Science and Technology Facilities Council (STFC), through a Doctoral Training grant - ST/T506011/1 and grants ST/R000425/1 and ST/V000934/1. 
C.O.G. Waterfall and S. Dalla acknowledge support from NERC grant NE/V002864/1.

This work was performed using resources provided by the Cambridge Service for Data Driven Discovery (CSD3) operated by the University of Cambridge Research Computing Service (www.csd3.cam.ac.uk), provided by Dell EMC and Intel using Tier-2 funding from the Engineering and Physical Sciences Research Council (capital grant EP/P020259/1), and DiRAC funding from the Science and Technology Facilities Council (www.dirac.ac.uk).
\end{acknowledgements}

\end{document}